
%

\hsize=6.0truein
\vsize=8.5truein
\voffset=0.25truein
\hoffset=0.1875truein
\tolerance=1000
\hyphenpenalty=500
\def\monthintext{\ifcase\month\or January\or February\or
   March\or April\or May\or June\or July\or August\or
   September\or October\or November\or December\fi}


\font\tenrm=cmr10 scaled \magstep1   \font\tenbf=cmbx10 scaled \magstep1
\font\sevenrm=cmr7 scaled \magstep1  
\font\fiverm=cmr5 scaled \magstep1   

\font\teni=cmmi10 scaled \magstep1   \font\tensy=cmsy10 scaled \magstep1
\font\seveni=cmmi7 scaled \magstep1  \font\sevensy=cmsy7 scaled \magstep1
\font\fivei=cmmi5 scaled \magstep1   \font\fivesy=cmsy5 scaled \magstep1

\font\tentt=cmtt10 scaled \magstep1
\font\tenit=cmti10 scaled \magstep1
\font\tensl=cmsl10 scaled \magstep1

\def\twelvepoint{\def\rm{\fam0\tenrm}
   \textfont0=\tenrm \scriptfont0=\sevenrm \scriptscriptfont0=\fiverm
   \textfont1=\teni  \scriptfont1=\seveni  \scriptscriptfont1=\fivei
   \textfont2=\tensy \scriptfont2=\sevensy \scriptscriptfont2=\fivesy
   \textfont\itfam=\tenit \def\it{\fam\itfam\tenit}
   \textfont\ttfam=\tentt \def\tt{\fam\ttfam\tentt}
   \textfont\bffam=\tenbf \def\bf{\fam\bffam\tenbf}
   \textfont\slfam=\tensl \def\sl{\fam\slfam\tensl} \rm
   \hfuzz=1pt\vfuzz=1pt
   \setbox\strutbox=\hbox{\vrule height 10.2pt depth 4.2pt width 0pt}
   \parindent=24pt\parskip=1.2pt plus 1.2pt
   \topskip=12pt\maxdepth=4.8pt\jot=3.6pt
   \normalbaselineskip=14.4pt\normallineskip=1.2pt
   \normallineskiplimit=0pt\normalbaselines
   \abovedisplayskip=13pt plus 3.6pt minus 5.8pt
   \belowdisplayskip=13pt plus 3.6pt minus 5.8pt
   \abovedisplayshortskip=-1.4pt plus 3.6pt
   \belowdisplayshortskip=13pt plus 3.6pt minus 3.6pt
   \topskip=12pt \splittopskip=12pt
   \scriptspace=0.6pt\nulldelimiterspace=1.44pt\delimitershortfall=6pt
   \thinmuskip=3.6mu\medmuskip=3.6mu plus 1.2mu minus 1.2mu
   \thickmuskip=4mu plus 2mu minus 1mu
   \smallskipamount=3.6pt plus 1.2pt minus 1.2pt
   \medskipamount=7.2pt plus 2.4pt minus 2.4pt
   \bigskipamount=14.4pt plus 4.8pt minus 4.8pt}

\twelvepoint



\font\titlerm=cmr10 scaled \magstep3
\font\titlerms=cmr10 scaled \magstep1 
\font\titlei=cmmi10 scaled \magstep3  
\font\titleis=cmmi10 scaled \magstep1 
\font\titlesy=cmsy10 scaled \magstep3 	
\font\titlesys=cmsy10 scaled \magstep1  
\font\titleit=cmti10 scaled \magstep3	
\skewchar\titlei='177 \skewchar\titleis='177 
\skewchar\titlesy='60 \skewchar\titlesys='60 

\def\titlefont{\def\rm{\fam0\titlerm}
   \textfont0=\titlerm \scriptfont0=\titlerms 
   \textfont1=\titlei  \scriptfont1=\titleis  
   \textfont2=\titlesy \scriptfont2=\titlesys 
   \textfont\itfam=\titleit \def\it{\fam\itfam\titleit} \rm}


\def\preprint#1{\baselineskip=19pt plus 0.2pt minus 0.2pt \pageno=0
   \begingroup
   \nopagenumbers\parindent=0pt\baselineskip=14.4pt\rightline{#1}}
\def\title#1{
   \vskip 0.9in plus 0.45in
   \centerline{\titlefont #1}}
\def\secondtitle#1{}
\def\author#1#2#3{\vskip 0.9in plus 0.45in
   \centerline{{\bf #1}\myfoot{#2}{#3}}\vskip 0.12in plus 0.02in}
\def\secondauthor#1#2#3{}
\def\addressline#1{\centerline{#1}}
\def\abstract{\vskip 0.7in plus 0.35in
	\centerline{\bf Abstract}
	\smallskip}
\def\finishtitlepage#1{\vskip 0.8in plus 0.4in
   \leftline{#1}\supereject\endgroup}

\def\date#1{\finishtitlepage{#1}}

\def\nolabels{\def\eqnlabel##1{}\def\eqlabel##1{}\def\figlabel##1{}%
	\def\reflabel##1{}}
\def\writelabels{\def\eqnlabel##1{%
	{\escapechar=` \hfill\rlap{\hskip.11in\string##1}}}%
	\def\eqlabel##1{{\escapechar=` \rlap{\hskip.11in\string##1}}}%
	\def\figlabel##1{\noexpand\llap{\string\string\string##1\hskip.66in}}%
	\def\reflabel##1{\noexpand\llap{\string\string\string##1\hskip.37in}}}
\nolabels


\global\newcount\secno \global\secno=0
\global\newcount\meqno \global\meqno=1

\def\newsec#1{\global\advance\secno by1
   \xdef\secsym{\the\secno.}
   \global\meqno=1\bigbreak\medskip
   \noindent{\bf\the\secno. #1}\par\nobreak\smallskip\nobreak\noindent}
\xdef\secsym{}

\def\appendix#1#2{\global\meqno=1\xdef\secsym{\hbox{#1.}}\bigbreak\medskip
\noindent{\bf Appendix #1. #2}\par\nobreak\smallskip\nobreak\noindent}

\def\acknowledgements{\bigbreak\medskip\centerline{\bf
   Acknowledgements}\par\nobreak\smallskip\nobreak\noindent}


\def\eqnn#1{\xdef #1{(\secsym\the\meqno)}%
	\global\advance\meqno by1\eqnlabel#1}
\def\eqna#1{\xdef #1##1{\hbox{$(\secsym\the\meqno##1)$}}%
	\global\advance\meqno by1\eqnlabel{#1$\{\}$}}
\def\eqn#1#2{\xdef #1{(\secsym\the\meqno)}\global\advance\meqno by1%
	$$#2\eqno#1\eqlabel#1$$}


\def\myfoot#1#2{{\baselineskip=14.4pt plus 0.3pt\footnote{#1}{#2}}}
\global\newcount\ftno \global\ftno=1
\def\foot#1{{\baselineskip=14.4pt plus 0.3pt\footnote{$^{\the\ftno}$}{#1}}%
	\global\advance\ftno by1}


\global\newcount\refno \global\refno=1
\newwrite\rfile

\def\ref{[\the\refno]\nref}
\def\nref#1{\xdef#1{[\the\refno]}\ifnum\refno=1\immediate
	\openout\rfile=refs.tmp\fi\global\advance\refno by1\chardef\wfile=\rfile
	\immediate\write\rfile{\noexpand\item{#1\ }\reflabel{#1}\pctsign}\findarg}
\def\findarg#1#{\begingroup\obeylines\newlinechar=`\^^M\passarg}
	{\obeylines\gdef\passarg#1{\writeline\relax #1^^M\hbox{}^^M}%
	\gdef\writeline#1^^M{\expandafter\toks0\expandafter{\striprelax #1}%
	\edef\next{\the\toks0}\ifx\next\null\let\next=\endgroup\else\ifx\next\empty%

 \else\immediate\write\wfile{\the\toks0}\fi\let\next=\writeline\fi\next\relax}}
	{\catcode`\%=12\xdef\pctsign{
\def\striprelax#1{}

\def\semi{;\hfil\break}
\def\addref#1{\immediate\write\rfile{\noexpand\item{}#1}} 

\def\listrefs{\vfill\eject\immediate\closeout\rfile
   \centerline{{\bf References}}\bigskip{\frenchspacing%
   \catcode`\@=11\escapechar=` %
   \input refs.tmp\vfill\eject}\nonfrenchspacing}

\def\startrefs#1{\immediate\openout\rfile=refs.tmp\refno=#1}


\global\newcount\figno \global\figno=1
\newwrite\ffile
\def\fig{\the\figno\nfig}
\def\nfig#1{\xdef#1{\the\figno}\ifnum\figno=1\immediate
	\openout\ffile=figs.tmp\fi\global\advance\figno by1\chardef\wfile=\ffile
	\immediate\write\ffile{\medskip\noexpand\item{Fig.\ #1:\ }%
	\figlabel{#1}\pctsign}\findarg}

\def\listfigs{\vfill\eject\immediate\closeout\ffile{\parindent48pt
	\baselineskip16.8pt\centerline{{\bf Figure Captions}}\medskip
	\escapechar=` \input figs.tmp\vfill\eject}}


\def\letter{\raggedright\parindent=0pt}
\def\endmode{}
\def\longindent{\parindent=3.25truein\obeylines\parskip=0pt}
\def\letterhead{\null\vfil\begingroup
   \parindent=3.25truein\obeylines
   \def\endmode{\medskip\endgroup}}

\def\sendingaddress{\endmode\begingroup
   \parindent=0pt\obeylines\def\endmode{\medskip\endgroup}}

\def\salutation{\endmode\begingroup
   \parindent=0pt\obeylines\def\endmode{\medskip\endgroup}}

\def\body{\endmode\begingroup\parskip=\smallskipamount
   \def\endmode{\medskip\endgroup}}

\def\closing{\endmode\begingroup\longindent
   \def\endmode{\endgroup}}

\def\signed{\endmode\begingroup\longindent\vskip0.8truein
   \def\endmode{\endgroup}}

\def\endofletter{\endmode \ifnum\pageno=1 \nopagenumbers\fi
	\vfil\vfil\eject\end}


\def\noblackbox{\overfullrule=0pt}
\def\inv{^{\raise.18ex\hbox{${\scriptscriptstyle -}$}\kern-.06em 1}}
\def\dup{^{\vphantom{1}}}
\def\Dsl{\,\raise.18ex\hbox{/}\mkern-16.2mu D} 
\def\dsl{\raise.18ex\hbox{/}\kern-.68em\partial}
\def\slash#1{\raise.18ex\hbox{/}\kern-.68em #1}
\def\lspace{}
\def\lbspace{}
\def\boxeqn#1{\vcenter{\vbox{\hrule\hbox{\vrule\kern3.6pt\vbox{\kern3.6pt
	\hbox{${\displaystyle #1}$}\kern3.6pt}\kern3.6pt\vrule}\hrule}}}
\def\mbox#1#2{\vcenter{\hrule \hbox{\vrule height#2.4in
	\kern#1.2in \vrule} \hrule}}  
\def\bar{\overline}
\def\e#1{{\rm e}^{\textstyle#1}}
\def\del{\partial}
\def\curly#1{{\hbox{{$\cal #1$}}}}
\def\curlyD{\hbox{{$\cal D$}}}
\def\curlyL{\hbox{{$\cal L$}}}
\def\vev#1{\langle #1 \rangle}
\def\psibar{\overline\psi}
\def\lform{\hbox{$\sqcup$}\llap{\hbox{$\sqcap$}}}
\def\darr#1{\raise1.8ex\hbox{$\leftrightarrow$}\mkern-19.8mu #1}
\def\half{{\textstyle{1\over2}}} 
\def\roughly#1{\ \lower1.5ex\hbox{$\sim$}\mkern-22.8mu #1\,}
\def\MSbar{$\bar{{\rm MS}}$}
\hyphenation{di-men-sion di-men-sion-al di-men-sion-al-ly}

\def\R{{\cal R}}
\def\frac#1#2{{#1\over#2}}
\def\hf{\half}
\def\nonp{non-perturbative}
\def\hmm{hermitian matrix model}
\def\sqs{stochastic quantization scheme}
\def\smm{supersymmetric matrix model}
\def\rline{{\rm I}\!{\rm R}}
\def\integ#1#2#3{\int_{#1}^{#2}\!\!\! d#3\ }
\preprint{\vbox{\rightline{SHEP 90/91--29}
\vskip2pt\rightline{NBI-HE-91-27}}}
\title{\vbox{\centerline{Stochastic Quantization vs. KdV Flows}
\vskip2pt\centerline{in}
\vskip2pt\centerline{2D Quantum Gravity}}}
\author{J. Ambj\o rn,$^1$ C.V. Johnson$^2$ and T.R. Morris$^2$}{}{}
\addressline{\it ${}^1$Niels Bohr Institute}
\addressline{\it Blegdamsvej 17, DK-2100 Copenhagen \O ,  Denmark}
\addressline{and}
\addressline{\it ${}^2$Physics Department}
\addressline{\it The University of Southampton}
\addressline{\it SO9 5NH, U.K.}

\abstract

We consider the stochastic quantization scheme for a non-perturbative
stabilization of 2D quantum gravity and prove that it does not
satisfy the KdV flow equations. It therefore differs from a recently
suggested matrix model which allows real solutions to the KdV equations.
The behaviour of  the Fermi energy, the free energy and macroscopic loops
in the stochastic quantization scheme are elucidated.

\date{August 1991}

\def\O{{\cal O}}
\newsec{Introduction}
There are now at least two proposals for a non-perturbatively
stable formulation of two dimensional quantum gravity. On the
one hand there is the ``5th time''
method\ref\Greensite{J. Greensite and M.
 Halpern,
Nucl. Phys. {\bf B242} (1984)
167.}
rediscovered in the context of
 Parisi-Sourlas reduction of a one dimensional
 supersymmetric matrix model\ref\Mari{E.Marinari and
G.Parisi, Phys. Lett. {\bf 240B} (1990) 375.}
a.k.a. stochastic quantization\foot{Other suggestions based on
stochastic quantization are to be found in ref.\ref\Marii{
E.Marinari and G.Parisi, Phys. Lett. {\bf 247B} (1990)
 537.}.}\ref\stoch{J.Ambj\o rn,
  J.Greensite and S.Varsted, Phys. Lett. {\bf
249B} (1990) 411.}\ref\migk{M.Karliner
and A.Migdal, Mod. Phys. Lett. {\bf 5A}
 (1990) 2565.}\ref\janjeff{J.Ambj\o rn and
J.Greensite, Phys. Lett. {\bf B254} (1991) 66.}\ref\spanish{
J.Gonz\'alez and M.A.H. Vozmediano, Instituto de Estructura de la
Materia preprint IEM-FT-35/90.}
and on the
other hand a definition\ref\mi{S.Dalley,
C.Johnson and T.Morris, Southampton
 Preprint SHEP 90/91--16.}\ref\mii{S.Dalley, C.Johnson
and T.Morris, Southampton Preprint SHEP 90/91--28.}
 following from \nonp\
KdV flows\ref\gm{D.J. Gross and A.A. Migdal,
 Nucl. Phys. {\bf B340} (1990)
 333.}\ref\KdV{T.Banks, M.Douglas, N.Seiberg and
S.Shenker, Phys. Lett. {\bf 238B} (1990) 279.}. Both proposals
agree\mi\Mari\
with the perturbative (i.e. genus by genus) results from
hermitian matrix models\ref\lots{E.Br\'{e}zin and
 V.Kazakov, Phys. Lett. {\bf
 B236} (1990) 144\semi
M.Douglas and S.Shenker, Nucl. Phys. {\bf B335} (1990) 135\semi
D.Gross and A.A.Migdal, Phys. Rev. Lett. {\bf 64} (1990) 127.}
as indeed they must, while also providing (apparently) a
physically sensible non-perturbative extension.

By ``physically
sensible'' we mean at least that the string susceptibility
and correlations of all local operators are real and free
of singularities (on the real axis). This has been shown to be
true in the KdV flow method\mi\mii , and in the stochastic
quantization method can be expected from the construction;
Recall that it is not true of the full \nonp\
result derived from hermitian matrix models
\ref\davi{F.David, Mod. Phys. Lett. {\bf A5}
(1990) 1019.}\ref\davii{F.David Nucl. Phys. {\bf B348} (1991)
 507.}. We also require that
macroscopic loop expectation values be physically sensible.
For the KdV flow method this has been checked in ref.\mii .
Part of the purpose of this paper will be to demonstrate
similarly sensible results for the stochastic quantization
method.\foot{Our interest in this problem was triggered by some remarks
in
ref.\ref\daviii{F.David,
 Notes added to talk given at the Cargese Workshop
 ``Random Surfaces, Quantum Gravity and Strings", May
28-June 1, 1990. } and it seems that our conclusions differ from the
the ones presented there.}

Na\"\i vely one might expect that 2D quantum gravity has a unique
\nonp\ extension, and hence that the two proposals are
equivalent.
This is not the case. We demonstrate this below,
both numerically and analytically.
It follows unfortunately that at present the question of the
``correct'' \nonp\ extension becomes  a philosophical
 one. The problem is of course that no general
principle of non-perturbative gravity has yet been defined,
and since both methods have the same perturbative expansion,
it is hard to promote one method over another if they have no
obvious flaws.

The proof that the KdV flows are violated non-perturbatively
in the stochastic quantization scheme will be  as
follows. The KdV flows induced by the operators $\O_k$ are
given by\KdV :
\eqn\flow{{\partial\rho\over\partial t_k} =
R'_{k+1}[\rho]
}
Here the $t_k$'s are couplings
conjugate
to the operators $\O_k$ and $\rho$ is the string
susceptibility related to the free energy $\Gamma$ by
$\Gamma'' =-\rho$. $\Gamma$ is given in terms of the
partition function as $Z=\exp(\Gamma)$.
Primes refer to
 differentiation with respect to $\mu$, the
cosmological constant. (The string coupling has been absorbed into
$\mu$ and the couplings $t_k$). The
$R_k$'s are the Gelfand--Dikii differential
 polynomials\ref\geldik{
I.R.Gelfand and L.A.Dikii, Russian Math. Surveys {\bf 30} (1975) 77.}.

If we assume that no further dimensionful parameter arises
at the non-perturbative level, then
these equations \flow\ are satisfied \nonp ly {\sl if and only
if} $\rho$ satisfies the string eqn.\mii :
\eqn\smiley{
0=\rho{\cal R}^2-\frac{1}{2} {\cal R} {\cal R}^{\prime\prime}
+\frac{1}{4} \left({\cal R}^{\prime}\right)^2
}
where for the full massive theory
\eqn\kdvtype{
{\cal R}
=\sum^\infty_{k=0} (k+\hf)\  t_k R_k[\rho] \ -\mu\ \ .}
For the purposes of this paper however, we will only be
interested in pure gravity in which case,
without loss of generality,
we take only $t_2\ne 0$ and
\eqn\oldnews{
\R=\rho^2-\rho''/3-\mu\ \ .
}
Now for negative cosmological constant $\mu$, if $\rho$ satisfies
eqn. \smiley\ and has an asymptotic expansion then it follows
from \oldnews\ and \smiley\ that the spherical
 contribution\foot{up to analytic terms in $\mu$ which are always
discarded}\
must vanish $\rho_0=0$ or be imaginary $\rho_0=\pm i |\mu|^{1/2}$.
(These are the roots of the algebraic equation obtained by deleting
all derivative terms in \smiley ). For negative $\mu$ an
asymptotic expansion for $\rho$ does indeed exist
in the supersymmetric
1D matrix model since this expansion
just corresponds to the WKB expansion.
Because it is real the leading term is of
the form $\rho_0=\kappa \sqrt{|\mu|}$ for some real constant
$\kappa$, by scaling arguments. We will prove
that $\kappa$ is different from zero.
It follows that the string susceptibility
calculated in the stochastic quantization method
cannot satisfy \smiley\ and hence violates eqns. \flow\
\nonp ly.

We note in passing that it is in fact natural to introduce
a new non-perturbative parameter $\sigma$ into \smiley\
corresponding to the position of the eigenvalue space
boundary\ref\talk{S.Dalley, C.Johnson and T.Morris, Southampton
Preprint SHEP-90/91-35, talk given at Barcelona Workshop
``Random Surfaces and 2D Quantum Gravity'', 10-14 June 1991
to appear in Nucl. Phys. {\bf B} Proc. Suppl.\semi
C.Johnson, T.Morris and B.Spence, in preparation.}.
It corresponds physically to a world-sheet boundary cosmological
constant. In this case \smiley\ takes the form
$
0=(\rho-\sigma)
{\cal R}^2-\frac{1}{2} {\cal R} {\cal R}^{\prime\prime}
+\frac{1}{4} \left({\cal R}^{\prime}\right)^2
$.
It is equivalent to \smiley\ under the finite `gauge'
transformation $\exp\{-\sigma L_{-1}\}$ and so does
not alter our conclusions. ($\epsilon L_{-1}$ generates
the KdV galilean transformation: $\delta t_k =
\epsilon (k+3/2) t_{k+1},\ \ \delta\rho=\delta\sigma=
\epsilon$).

It may seem a touch paradoxical that \nonp\ information
can be deduced from perturbative arguments in the regime
$\mu<0$. We would therefore like to emphasise that the
asymptotic results for $\mu<0$ in both schemes should
properly be regarded as \nonp . From the physical point
of view both schemes derive their connection to pure
2D quantum gravity/0D string theory indirectly through
exact agreement with the $\mu>0$ \hmm\
perturbative results, the latter having a direct
representation in terms of regularised surfaces through
the dual of Feynman diagrams. Neither scheme has  a
{\sl direct} world-sheet interpretation. (This is
discussed further for the complex matrix model
representation of the KdV flow method in ref.\mi ).
When $\mu<0$ both schemes lose their
interpretation in terms of surfaces since,
if for no other reason, they now disagree with the
\hmm . Actually the \hmm\ has
no interpretation in terms of surfaces
here either because the $\mu<0$ region is outside the
radius of convergence of its large $N$ Feynman diagram
expansion.
Indeed mathematically the {\sl perturbative}
results for $\mu<0$ are separated from those for
$\mu>0$ by non-smooth behaviour at $\mu=0$.
 The spherical free energy suffers
a 3rd order phase transition; In the \sqs\ for
$\mu>0$ supersymmetry is preserved to all orders
of perturbation theory, while for $\mu<0$ supersymmetry
is broken even on the sphere. In the representation
of the KdV flow scheme in terms of an eigenvalue
spectrum $\lambda\in\rline_+$ the classical
eigenvalue distribution hits the ``wall'' at $\lambda=0$
as $\mu\to0^+$ \mi\mii (more generally\talk\ $\lambda
=\sigma$ as $\mu\to(\sigma^2)^+$).

The rest of this paper is organised as follows:
Section 2 derives the spherical string susceptibility
for $\mu<0$ by calculating the one point function.
The role of the Fermi energy and the free energy in the supersymmetric
matrix model is clarified. Macroscopic loops on the sphere
are discussed in section 3 and a non-polynomial
0D \hmm\ potential that produces the same results is derived.
A numerical comparison
of the one-point function in both schemes is made in
sect. 4, and in sect. 5 we draw our conclusions.

\newsec{The Sphere for Negative Cosmological Constant.}

Recall that the  stochastic quantization scheme for the simplest
matrix models with cubic  potentials results in
replacing expectation values defined in the hermitian matrix model
\eqn\exvalue{
<{\cal O}(\phi)> =
 {1 \over Z} \int d\phi \; {\cal O}(\phi) \exp (-N \; {\rm Tr}
\; V(\phi))
}
where $\phi$ is an $N\times N$ hermitian matrix, with expectation values
of the same operator in the ground state of $N$ non-interacting
fermions which have the single particle Fokker-Planck (FP) hamiltonian
\eqn\fphamilton{
H_{FP} = N [ -{1 \over N^2}{d^2 \over d \lambda^2} + V_{FP} (\lambda)]
}
where
\eqn\fppotential{
V_{FP}(\lambda) = {1 \over 4} (V'(\lambda))^2-{1\over 2}V''(\lambda)
}
Whereas the simplest
potentials $V$ that give ordinary $k=2$ critical behaviour
are unbounded from below, the Fokker-Planck potential $V_{FP}$ is not,
and this is the reason the method leads to non-perturbative stabilization.
In fact the expectation value of any observable, which in the
hermitian matrix model will be given by \exvalue\ , is here defined by
\eqn\exfpvalue{
<{\cal O}(\phi)>_{FP} = \int\!d\phi \; \Psi^2_0(\phi) {\cal O}(\phi),
}
where $\Psi_0(\phi)$ is the groundstate wavefunction of the $N$-particle
fermionic system.  In the following we will restrict ourselves to the
spherical approximation and  to observables of the form
${\cal O}(\phi)= {\rm Tr}\;f(\phi)$. The spherical approximation is in
the hamiltonian language equivalent to the WKB approximation
and the expectation value of the observables ${\rm Tr}\; f(\phi)$ will
be given by:
\eqn\exwkb{
{1\over N} < {\rm Tr}\; f(\phi)> = {1\over \pi}
\integ{\lambda_l}{\lambda_r}{\lambda} \;\sqrt{E_F-
V_{FP}(\lambda)} \; f(\lambda).
}
Here $\lambda_r$ and $\lambda_l$ denote the left- and right turning
points, and $E_F$ denotes the Fermi energy of the system, i.e. the $N$'th
energy level, scaled by $N^{-1}$.
 The WKB condition for the $N$'th energy level of the
hamiltonian \fphamilton\ is
\eqn\wkbenergy{
1 = {1\over \pi}
\integ{\lambda_l}{\lambda_r}{\lambda} \;\sqrt{E_F-V_{FP}(\lambda)}
}

To be
explicit let us consider the simplest potential $V$ in the hermitian
matrix model which leads to $k=2$ critical behaviour:
\eqn\simple{
V(\phi)= g\phi - \phi^3/3
}
The corresponding FP-potential is
\eqn\FP{
V_{FP}(\lambda;g) = {1 \over 4} (g-\lambda^2)^2 +\lambda
}
The critical coupling  in the matrix model is $g_c = 3/2^{2/3}$ and
the endpoint of the eigenvalues distribution is
$\lambda_c= \sqrt{g_c/3}= 1/2^{1/3}$ where $V(\lambda_c;g_c)=g^2_c/3=
3\lambda^4_c$ and the derivatives
with respect
 to $\lambda$ satisfy: $ V'(\lambda_c,:g_c)=V''(\lambda_c;g_c)=0$.

If we introduce the scaled variables:
\eqn\scale{
 g -g_c = g_c a^2 \mu,~~~~~\lambda-\lambda_c = \lambda_c  a y,~~~~
a= 1/N^{2/5}
}
we can write
$$\eqalignno{
V_{FP} (y;\mu)& = V_{FP}(0;\mu )+\lambda_c^4 a^3 v_{fp}(y;\mu) & \cr
v_{fp}(y;\mu ) & = -3\mu y + y^3+a\bigr[ -{3 \over 2} \mu\,y^2+
{1\over 4} \, y^4\bigl]    & \cr
}$$
where $V_{FP}(0;\mu)$ is a second order polynomial in $g-g_c=g_ca^2\mu$.
As in \stoch\ the FP-hamiltonian can be written
$$\eqalignno{
H_{FP}& = N [ V_{FP} (0;\mu ) +\lambda_c^4 a^3  h_{fp} ]  &\cr
h_{fp}& =  -4{d^2 \over dy^2} - v_{fp}(y;\mu)          & \cr
}$$
and if we introduce the scaled Fermi energy $e_f$ by
\eqn\energy{
E_F= V_{FP}(0;\mu )+ \lambda_c^4 a^3 \; e_f
}
the equation \wkbenergy\ for the Fermi energy can be written as
\eqn\normal{
N= {1 \over 2\pi} \integ{y_l}{y_r}{y}\; \sqrt{ e_f - v_{fp}(y;\mu)}
}

While the original hermitian matrix model even at the spherical level
is defined only for $g > g_c$, i.e. for $\mu > 0$, the FP-potential
is defined for all real $\mu$.
Since $E_F$ in the spherical limit can be a function of the coupling
constant $g$ only (i.e. $g_c$ and $g -g_c$), we conclude that
\eqn\ef{
e_f= |\mu|^{3/2} f^{(\pm)}(a\sqrt{ |\mu |}).
}
where $f^{(\pm)}(\cdot)$ might be different functions for $\mu >0$ and
$\mu < 0$. Let us assume that the $f$'s have the following expansions
\eqn\fexp{
f^{(\pm)}(x) = f_0^{(\pm)} + O(x),
}
where the constants $f_0^{(\pm)}$ are to be determined from \normal .
By differentiating  \normal\ with respect to $|\mu|$ and taking the limit
$a \to  0$ one gets the following equations for$f_0^{(\pm)}$:
\eqn\feqn{
0= \lim_{a\to 0} \;\;\integ{y_l}{y_r}{y} \;
{{{1\over 2} \sqrt{|\mu |} f_0^{(\pm)} \pm (y+ {1 \over 2} a y^2)} \over
\sqrt{ f_0^{(\pm)}|\mu|^{3/2}-v_{fp}(y;\mu)}} \; .
}
The integration limits, the turning points, before taking
$a \to 0$ are given by
\eqn\limits{
y_r = \sqrt{|\mu |}z_r(f_0^{(\pm)}) + O(a),~~~~~~
y_l= -{4 \over a} - {3 \over 4}\mu a + O(a^2).
}
Here $\sqrt{|\mu |}z_r$ denotes
the right turning point solution in the limit
$a \to 0$. We have
\eqn\zr{
f_0^{(\pm)}\pm 3z_r -z_r^3 = 0.
}
Subtracting from $\pm$\feqn\ the same equation with $\mu=0$, and
then changing variables as $y\to \sqrt{|\mu|} z$ we obtain the
convergent integrals:
\eqn\feqns{
 \integ{-\infty}{z_r}{z} {{1\over 2}f_0^{(\pm)}\over \sqrt{f_0^{(\pm)}
             \pm 3z-z^3}} =
\mp \integ{-\infty}{z_r}{z} [ \;
{z \over \sqrt{f_0^{(\pm)}\pm 3z-z^3}}-{z\over \sqrt{|z|^3 }} \;]\; \pm
\integ{z_r}{0}{z} \; {z\over \sqrt{|z |^3}}
}
The solution of \feqns\ is well known for $\mu > 0$
due to the inherent supersymmetry of the FP-hamiltonian \Mari ,
which is unbroken for $\mu > 0$, and is given by
\eqn\solplus{
f_0^{(+)}= -2~~~~~({\rm and}~~~z_r=-2)\;\; .
}
For $\mu < 0$ it is readily seen that $f_0^{(-)} > 0$ and $z_r > 0$,
but the integrals
can not be expressed as elementary integrals. Numerical integration gives
\eqn\solminus{
f_0^{(-)}\approx 8.0   ~~~~~~({\rm and}~~~z_r\approx 1.52)\;\; .
}

Let us now turn to the observables of the theory. The simplest and
most fundamental ones in the
hermitian matrix model are the  puncture operator and the susceptibility.
In the case of the
potential \simple\ the `bare' puncture operator and the susceptibitity
can be defined as
 follows
\eqn\puncture{
<{\cal O}_0>= -{1\over N^2} \;\Gamma' (g) = < {1\over N} \; {\rm Tr}\,
\phi>.
}
\eqn\suscep{
\rho(g)  = -{d <{\cal O}_0> \over  dg} = {1 \over N^2} \Gamma''(g)
}
In \puncture\ $\Gamma$ denotes the free energy of the
hermitian matrix model.
The FP-version of \puncture\ can be found from  \exfpvalue\ , or in the
spherical approximation  from \exwkb\ ~
\eqn\fppunct{
<{\cal O}_0>_{FP} = {1\over \pi}
\integ{\lambda_l}{\lambda_r}{\lambda} \; \lambda\;
\sqrt{E_F -V_{FP}(\lambda)} \;  = \;
\lambda_c ( 1+ {a^{7/2}\over 2\pi}
\integ{y_l}{y_r}{y} \; y\;\sqrt{ e_f - v_{fp}(y)} \; \;).
}
and we get
\eqn\rhofp{
\rho_{FP} (\mu )=
{\lambda_c \over g_c} \; {a^{3/2} \over 4\pi}
\integ{y_l}{y_r}{y} \;
{y ( e_f' + 3 (y+ {1\over 2} a y^2) ) \over   \sqrt{ e_f - v_{fp}(y)} }
}
Inserting the integration limits \limits\  and subtracting, as in \feqns ,
the expression for $\mu =0$, leads to the following result :
\eqn\rhosub{
\rho_{FP}(\mu)-\rho_{FP}
(0)={\lambda_c \over 2g_c} \;\left( c_1 e_f'(\mu) a +
 O((\sqrt{|\mu |} a)^{3/2})\right)
}
where
$$
c_1= {\sqrt{a}\over2\pi}
\integ{-4/a}{0}{y} \; {y \over \sqrt{-y^3-{a \over 4} y^4}} =-1
$$
If we recall that $g-g_c = g_c a^2\mu$, the leading non-analytic power
of $g-g_c$ is given by the term involving the Fermi energy.
Since it is different from zero for both positive and negative
cosmological constants, according to \solplus\ and \solminus\
 we conclude that the \smiley\ must be violated even at the spherical
level by the stochastic regularization.

If we integrate \rhosub\ with respect to $g-g_c$ we get
\eqn\omu{
< {\cal O}_0 >_{FP} = {\lambda_c\over2}\left(
 c_0 + c_2 \mu a^2 - e_f a^3 + O((\sqrt{|\mu|}a)^{7/2})\right)
}
and from \rhosub\ and \omu\ it is clear that the leading non-analyticity
in  $\mu$ (or $g-g_c$) is entirely contained in the Fermi energy
$e_{f}$ which  therefore can be considered as the puncture
operator\foot{This identification
 of the Fermi energy as the puncture operator
is only valid for the simplest potentials $V(\phi)$ which allow a
factorization of the ground state
 $\Psi_0$ in single particle wave functions
\ref\charlotte{J. Ambj\o rn and C.F. Kristjansen, to appear}.}.

Let us end this section with a discussion of the r\^{o}le of relations
like \puncture : $< {\cal O}_0 > = -\Gamma'(g) /N^2$ in the
context of stochastic quantization. Due to the hidden supersymmetry, which
is unbroken at the spherical level for $\mu > 0$ it is clear that
such a relation can not be true if we for $\Gamma$ use the energy of
the $N$-Fermi system, since the total energy of the $N$-Fermi
system is exactly zero for all $\mu > 0$.
However, it is possible to have a somewhat
similar formula in the stochastic scheme. By definition we have
from the second equality in \puncture\ that $< {\cal O}_0>_{FP} =
< {\rm Tr} \; \phi/N>_{FP}$ and we can get  this quantity from
the free energy $\Gamma_{FP}$ of the $N$-Fermi system if we add
a source term   $J\;{\rm Tr}\; \phi/N$ to the FP-hamiltonian. This term
explicitly breaks supersymmetry and $\Gamma_{FP}(J)$ will be different
from zero and we have, as noticed in \janjeff :
\eqn\jf{
< {\cal O}_0 >_{FP} =  {d \Gamma_{FP}(J,g) \over d J}|_{J=0}
}
The modified single particle hamiltonian will be
\eqn\mod{
h_{fp} = -4{d^2 \over dy^2} +v_{fp}(y,\mu ,j,a)
}
where $j$ is an appropriately scaled source term and
\eqn\vfmod{
v_{fp}(y,\mu , j,a) = -(3\mu+j)y+y^3 + \bigr[-{3\over 2} \mu ay^3+
{a \over 4}y^4\bigl]
}
It is now clear that taking  the limit $ a \to 0$ before calculating
expectation values
means that we are breaking supersymmetry in a way identical
to adding a source term $j$ since only the combination $3\mu+j$ appears
in \vfmod\ after $a$ is zero.
Of course the integrals defining the expectation values will
now be divergent
 since we have dropped the stabilizing term in the potential.
However, if one
 just introduces a cut-off $\Lambda \sim -1/a$ in the integrals
it can be shown
 that the leading non-analytical behaviour is still the same as
with the full potential. In this way one can convert \jf\ to an equation
similar to \puncture\ where the differentiation is with respect to
$\mu$ and where $\Gamma_{FP} (\mu)$ is the energy calculated from the
$N$-Fermion system with the unbounded potential $-3\mu y +y^3$, but with
a lower cut-off $\Lambda \sim -1/a$. All results in this section
can be derived using this technique.

\newsec{Macroscopic Loops on the Sphere.}

Motivated by ref.\daviii , we make some
remarks on the expectation value of a single macroscopic loop
in the \sqs . We only consider the spherical limit, however
as we have seen above the behaviour of the spherical limit
for $\mu <0$ is in a sense \nonp . It is not unreasonable
to expect the full \nonp\ result to have a similar
qualitative (but more smoothed out) behaviour.

{}From
identification through the (dual of the) Feynman diagram
expansion in the \hmm\ we know that the operator representing
the insertion of an $n$- polygonal loop is given by
\eqn\npolya{
<W(n)> = < {1\over N} \;{\rm Tr}\; \phi^n>
}
This class of observables belongs to the ones which can readily
be defined in the stochastic regularization scheme. In the
spherical approximation \exwkb\ we have
\eqn\npolyb{
<W(n)>= {1\over \pi} \integ{\lambda_r}{\lambda_l}{\lambda} \;
\sqrt{E_F-V_{FP}(\lambda)} \; \lambda^n.
}
For finite $n$ each observable clearly has  the  same
critical behaviour as the puncture operator. However
interesting behaviour arises if
one scales $n \to \infty$ as
\eqn\nl{
n = l/a.
}
In this way we can identify $l$ with a macroscopic length of the
loop, $a$ being identified with the "lattice" spacing. Introducing
the scaled variables \scale\ in \npolyb\ we get
\eqn\npolyc{
<W({l\over a})> = {\lambda_c^{l/a}\over N} \;\; {1 \over 2\pi}
\integ{y_r}{y_l}{y} \sqrt{e_f-v_{fp}(y)} \; (1+ay)^{l/a}.
}
In the scaling limit $(1+ay)^{l/a} \to e^{ly}$ and we get
\eqn\defw{
<W({l\over a})> ={\lambda_c^{l/ a}\over N}  \; <w(l)>;~~~~~~~~
<w(l)>\equiv  {1\over 2\pi}
\integ{y_r}{y_l}{y} \sqrt{e_f-v_{fp}(y)} \;\e{ly}.
}
This expression is perfectly well defined  for both positive and
negative $\mu$ as long as
 $l$ is positive. For $l$ negative \npolya -\npolyc\
are not well defined and this is reflected in the exponential divergence
of \defw\ for $a \to 0$. As is well known \davi
\ref\am{J. Ambj\o rn and Y. Makeenko, Mod. Phys. Lett. A5 (1990) 1753}
\ref\ajm{J. Ambj\o rn,
J. Jurkiewicz and Y. Makeenko, Phys.Lett. 251B (1990) 517}
the Laplace transform of $w(l)$ is the generator of one point functions
 and has very transparent analyticity properties. It is defined by
\eqn\F{
F(z) = \integ{0}{\infty}{l} \;
\e{-lz} <w(l)>\; = \; {1\over 2\pi}\integ{y_l}{y_r}{y}\;
{\sqrt{e_f-v_{fp}(y)}\over y-z}
}
{}From this integral representation it is clear that $F(z)$ is analytic
everywhere in the complex plane except along the cut $ z \in [y_l,y_r]$
and we can express $<w(l)>$ by the inverse laplace transform:
\eqn\w{
<w(l)>=\int_{-i\infty+r}^{i\infty+r}{dz\over2\pi i}\ \e{zl}
F(z)\ \ .
}
Here $r$ is to the right of any singularities of $F(z)$ in the
$z$ complex plane, i.e. to the right of $y_r$.
In fact \w\ makes sense even for negative $l$ and can be
considered as a definition of negative length loops. For $l$
negative we can clearly close the contour to the right in \w\ and
we get
\eqn\nothing{
<w(l)> = 0 ~~~~~{\rm for}~~~~ l < 0 \; .
}
This physically very reasonable result is true independently
of whether $\mu$ is positive or negative\foot{We should note that
our definition seems to differ from the one given in \daviii\ where
one gets an expectation value of $w(l)$ for $l < 0$.}.

Up to now the there has been no difference between negative and
positive cosmological constants $\mu$. However
 the cut-structures of
$2\pi u_{fp}(z) \equiv \sqrt{e_f-v_{fp}(z)}$
in the complex plane  are different.
The hidden supersymmetry, unbroken for $\mu >0$, ensure that in
this case $e_f -v_{fp}(z)$ has a double zero $y_0 = \sqrt{|\mu |}+ O(a)$
and in this case (recall that $y_l = -4/a + O(a)$):
\eqn\upos{
2\pi u_{fp}(z) = (y_0-z)\sqrt{(y_r-z)(z-y_l)a/4} \; .
}
However, for $\mu < 0$ the double zero $y_0$ splits
in two complex zeroes given approximately by
\eqn\cut{
z_{\pm} = -{y_r\over 2} \pm i { \sqrt{3} \over 2} \;
\sqrt{y_r^2+4|\mu|}
}
and
\eqn\uneg{
2\pi u_{fp}(z) = \sqrt{(z-z_{+})(z-z_{-})(y_r-z)(z-y_l)a/4}
}
has an additional  cut from $z_{+}$ to $z_{-}$.
{}From the integral representation \F\ it is clear that $F(z)$ falls off
as $1/z$ for $|z| \to \infty$, is analytic
except along the cut $[y_l,y_r]$ on the  real axis and that
\eqn\revlection{
F(y+i\varepsilon)-F(y-i\varepsilon)\Big|_{y\in(y_l,y_r)}
=2\pi  i u(y).
}

When $u(z)$ is given by \upos\ the cut of $u(z)$ coincide with the one of
$F(z)$ and $F(z)$ is uniquely determined as
\eqn\Fpos{
F(z) = {1 \over 2} V'(z) + i \pi u(z)
}
where ${1 \over 2} V'(z)$ is given by \simple\ and
determined uniquely by the requirement
that $F(z) = O(1/z)$ for $|z| \to\infty$.

For $\mu < 0$ the additional cut in $u(z)$ invalidates \Fpos , and one has
to subtract a function with this cut and which falls of as $1/z$. It is
uniquely determined as
\eqn\Xdef{
\half X'(z) = {1\over 2\pi i}\oint_{\cal C} dz' {u(z')\over z'-z}
}
where the contour ${\cal C}$ surrounds the new cut. $F$ is now given as
\eqn\Fneg{
F(z)={1\over 2} (V'(z)+X'(z)) + i\pi u(z).
}
{}From the classic work
\ref\biz{E. Brezin, C. Itzykson, G. Parisi and J.B Zuber,
Commun.math.Phys. 59, (1978) 35}\
of Brezin et al. it is known that
if the eigenvalue distribution of the 0D-matrix model
in the large $N$ limit is given
by $u(y)$, defined on an interval $[y_l,y_r]$, then the resolvent
$F(z)$, defined by\foot{We have the relation
$$R(w)\equiv {1 \over N} < {\rm Tr} \; {1\over \phi-w}> ={a^{3/2} \over
\lambda_c} \;F(z)\; ,~~~~~~w=\lambda_c(1+a z)
$$
where $R(w)$ denotes the resolvent in unscaled variables.}
$$
F(z) = \integ{y_l}{y_r}{y} {u(y) \over  y-z} \;\; ,
$$
is related to the potential $V_{eff}(\phi)$ of the matrix model by
\eqn\Feff{
F(y\pm i\epsilon )
={1 \over 2} {V'}_{eff}(y) \pm i\pi u(y),~~~~y \in [y_l,y_r].
}
{}From \Fpos\ and \Fneg\ we conclude that the 0D matrix model
effective potential which gives the same results as the FP-potential
in the spherical limit is given by
\eqn\effpot{
V_{eff} = V+\theta (-\mu) X \;\; .
}
Note that $X$ is non-polynomial.
It would be interesting to examine whether the full \sqs\
solution would also follow from $V_{eff}$.

\newsec{A Numerical Comparison.}
We have shown analytically in sects.1 and 2 that the \sqs\
is different from the KdV flow method, by comparing their
leading asymptotics as $\mu\to -\infty$. It is of interest
to compare them in the small $\mu$ regime too. We can do
this numerically using the one-point function for the
puncture operator, $<\O_0>$, calculated in ref.\janjeff\
in the \sqs\ and $<\O_0>$ calculated in the KdV flow
method in ref.\mii .

We refer the reader to these papers for more details on those
calculations. In ref.\janjeff\ the real and imaginary parts
of
the \nonp\ \hmm\ solution are also computed and plotted. We
have copied the five points for the \sqs\ (the circles)
and for the real part of the \hmm\ solution (the triangles)
onto
 fig.\fig\iii{The expectation
  value of the puncture operator with the
   sphere contribution removed; $z$ is proportional to the
cosmological constant. The triangles are the real part of the
\hmm\ result. The circles are values given by the \sqs , and the
unmarked line is the result of the KdV flow scheme.},
using the conventions of ref.\janjeff . The
remaining line
 in fig.1 is calculated in the KdV flow method\mii .
The errors in this line are negligable ($\sim 10^{-5}$).
Although the three graphs clearly differ, it is intriguing
that they are qualitatively so similar.

The translation from the conventions of ref.\mii\ to those
of ref.\janjeff\ is
as follows:
$$P-P_0 =8\sqrt{3} \mu^{-3/2} (<\O_0>-{2\over3}
\mu^{3/2})$$
 is the sphere subtracted one-point function,
up to factors as indicated, and is called ${\cal F}$
and ${\cal G}$ in ref.\janjeff . The horizontal coordinate
$z$ is proportional to the cosmological constant:
$z={1\over 4}(2\sqrt{3})^{2/5}\mu$.
These translations are derived using the recursion relations
of the orthogonal polynomial coefficients $\tilde{r}_n$
and $\tilde{a}_n$ (eqns.(26) and (27) of ref.\janjeff )
to calculate the spherical contribution of ref.\janjeff 's
definition of the one point function
$$<\O_0>\equiv <{1\over N}\; {\rm Tr}\; \varphi^2> = {1\over 3gN}
\sum^{N-1}_{n=0} \tilde{a}_n$$
in the contour rotated \hmm . Using the continuum limits
given there in eqns.(29-31)\foot{Eqn.(30) should read
 $a^2y_n=\cdots$}\
we find
\eqn\onep{
P={<{1\over N}\; {\rm Tr}\; \varphi^2>\over (1-g/g_c)^{3/2}}
=-8\sqrt{3} \integ1\infty{y} f(y)
}
where $f$ satisfies the Painlev\'e I equation in the
form
$${\sqrt{3}\over 48 z^{5/2}} {d^2f\over dy^2}+f^2 =y\ \ .$$
We use this and eqn.\oldnews\ to make the fundamental
identifications $f=-\rho(x)/\sqrt{\mu}$, $y=x/\mu$ and
$z={1\over4}(2\sqrt{3})^{2/5}\mu$ from which it follows
that \onep\ translates to
$$P={8\sqrt{3}\over \mu^{3/2}} \integ\mu\infty{x}
\rho(x)\ \ .$$

\vfill\eject

\newsec{Conclusions}

As shown above
the KdV flow symmetry of 2D quantum gravity is
broken \nonp ly by the real solution provided by \sqs .
Since another scheme exists
which preserves these flows \nonp ly and which allows for a
 real solution \mii , and since we have found no obvious unphysical
behaviour of physical observables in the two schemes (as discussed in
sec. 3),
one is left with ambiguities. On the one hand we have the
general  method of stochastic stabilization of bottomless actions
\Greensite , which works in any dimension. As already pointed
out in ref.\janjeff\ it disagrees with the stabilization via contour
rotation
\ref\hawk{
K. Gawedzi and A. Kupiainen, Nucl.
Phys. {\bf B257[FS14]} (1985) 474\semi
G. Gibbons, S. Hawking, and M. Perry,
Nucl. Phys. {\bf B138} (1978) 141.}.
 In 2d  gravity the
contour rotation method satisfy the KdV-flow symmetry, but leads to
unacceptable complex solutions. On the other hand the unique real solution
to \smiley\ is an alternative  candidate for a non-perturbative
definition of 2d gravity\mii . The preference of this definition
to the one of  stochastic stabilization would amount
to the declaration that
KdV-flows are a fundamental property of non-perturbative
2D quantum gravity. Keeping
in mind on the one hand
 the ultimate goals of understanding
higher dimensional gravity and string theory, and on the other,
the lack of understanding of the fundamental physics behind the
KdV-flow structure,
it seems to us rather that
the results obtained in this article highlight the
need for a better understanding of the basic principles underlying
non-perturbative quantum gravity.

\acknowledgements
TRM would like to thank the Neils Bohr Institute for hospitality
during the first stages of this work, and Simon Dalley for helpful
comments. CVJ thanks the S.E.R.C.
for financial support and JA thanks Charlotte F. Kristjansen and
Jens Lyng Petersen for helpful discussions.

\listrefs
\listfigs
\bye